\documentclass{Interspeech}
\usepackage{multirow,graphicx,cite,hyperref}

\interspeechcameraready

\title{Layer-Wise Analysis of Self-Supervised Representations for Age and Gender Classification in Children’s Speech}

\author[affiliation={1}]{Abhijit}{Sinha}
\author[affiliation={1}]{Harishankar}{Kumar}
\author[affiliation={1}]{Mohit}{Joshi}
\author[affiliation={1}]{Hemant Kumar}{Kathania}
\author[affiliation={2}]{~~~~~~~~~~~~~~~~~~~~~~~~~~~~Shrikanth}{Narayanan}
\author[affiliation={2}]{Sudarsana}{Reddy Kadiri}

\affiliation{Department of Electronics and Communication Engineering}{NIT Sikkim}{India}
\affiliation{Signal Analysis and Interpretation Lab (SAIL)}{University of Southern California}{USA}

\email{phec230023@nitsikkim.ac.in, b210046@nitsikkim.ac.in, b210055@nitsikkim.ac.in, hemant.ece@nitsikkim.ac.in, shri@usc.edu, skadiri@usc.edu}


\usepackage{comment}

\begin{document}

\maketitle

\begin{abstract}
Children’s speech presents challenges for age and gender classification due to high variability in pitch, articulation, and developmental traits. While self-supervised learning (SSL) models perform well on adult speech tasks, their ability to encode speaker traits in children remains underexplored. This paper presents a detailed layer-wise analysis of four Wav2Vec2 variants using the PFSTAR and CMU Kids datasets. Results show that early layers (1–7) capture speaker-specific cues more effectively than deeper layers, which increasingly focus on linguistic information. Applying PCA further improves classification, reducing redundancy and highlighting the most informative components. The Wav2Vec2-large-lv60 model achieves 97.14\% (age) and 98.20\% (gender) on CMU Kids; base-100h and large-lv60 models reach 86.05\% and 95.00\% on PFSTAR. These results reveal how speaker traits are structured across SSL model depth and support more targeted, adaptive strategies for child-aware speech interfaces.

\end{abstract}

\section{Introduction}
\label{sec:intro}

Accurate classification from children’s speech is critical for personalized learning and content filtering. However, children’s speech presents unique challenges due to rapid developmental changes in vocal tract morphology \cite{Vorperian2007VowelAS} and inconsistent articulation patterns \cite{koenig2008speech}. While self-supervised learning (SSL) models such as Wav2Vec2 have demonstrated impressive performance on adult speech tasks, their layer-wise efficacy on children’s speech remains under-explored. Data from the American Academy of Child and Adolescent Psychiatry ({\href{https://www.aacap.org/AACAP/Families_and_Youth/Facts_for_Families/FFF-Guide/Children-And-Watching-TV-054.aspx}{AACAP}}, May 2024) indicate that children ages 8-12 years spend approximately 4-6 hours per day engaging with screens, with adolescents often reaching up to 9 hours daily. Consequently, there is a need for robust age and gender classification systems that can reliably inform content filtering algorithms, ensuring that online experiences for children are both safe and appropriate.

Prior approaches often fine-tune SSL models for age estimation \cite{kang2023svldl} or employ complex fusion methods \cite{kitagishi23_interspeech}, both of which are computationally intensive. In contrast, our analysis shows that the initial layers of the SSL model, which capture essential acoustic information, generalize effectively to children’s speech without the need for additional fine-tuning. The acoustic properties of children’s voices, characterized by greater variability in pitch and formant frequencies \cite{study_formant}, as well as distinct pronunciation patterns compared to adults \cite{Vorperian2007VowelAS,lee1999acoustics}, introduce additional complexity to speech processing. Moreover, the scarcity of large annotated datasets \cite{yeung2018difficulties,claus2013survey} complicates the development of robust age and gender classification systems.

Previous research on children’s age and gender classification has utilized Gaussian mixture models, deep neural networks \cite{sanchez2021age,kwasny2021gender}, and temporal convolutional neural networks \cite{sanchez2022age} with features such as Mel-frequency cepstral coefficients (MFCCs) \cite{safavi2016speaker}. Recent studies have explored acoustic and prosodic features for classification tasks \cite{li2013automatic,kumari2024role}. Other works have investigated the effectiveness of SincNet over the ERB scale \cite{radha2024automatic}. Additionally, researchers have also examined the use of TDNNs and LSTM networks on raw waveforms to enhance classification accuracy \cite{sarma2020children, jia2019children}.

Recent advancements in SSL models such as Wav2Vec2 \cite{baevski2020wav2vec} have revolutionized speech processing by leveraging large-scale unlabeled data. These models have achieved state-of-the-art results in tasks like speaker recognition, language identification, emotion recognition, and pathology detection \cite{Pepino2021EmotionRF,sinha2024effect,grosz2022wav2vec2,Gao2023TwostageFO,9747379, wav2vec2_speaker,ssl_emotion,ANIDJAR2024124671,novoselov23_interspeech,tirronen2023utilizing,javanmardi2024pre,javanmardi2024exploring}. Yet, their application to children’s age and gender classification remains under-explored, with most studies relying on fine-tuning or multi-layer feature fusion that offer limited insights into layer-specific behavior.

In this study, we address these gaps by investigating Wav2Vec2 models with diverse pretraining and fine-tuning configurations for age and gender classification from children’s speech. We extract features from four Wav2Vec2 variants (base-100h, base-960h, large-960h-lv60, and large-960h-lv60-self) and train a Convolutional Neural Network (CNN) classifier on two benchmark datasets: PFSTAR \cite{russell2006pf} and CMU Kids \cite{eskenazi1997cmu}. Through a layer-wise analysis, we identify task-relevant representations, enabling more targeted feature selection for improved classification performance. Furthermore, we apply dimensionality reduction to assess the necessity of all extracted features. Our findings indicate that reducing the feature set improves classification accuracy, suggesting that not all SSL feature representations are necessary for accurate age and gender classification. 

Our approach addresses two primary research questions, which we explore through a rigorous experimental setup:

\begin{itemize}
    \item \textbf{How do different layers in SSL models encode age and gender cues?}    
    Our experiments reveal that early layers predominantly encode speaker-specific age and gender-related features, whereas deeper layers progressively abstract these details, focusing more on linguistic and phonetic representations.

    \item \textbf{Which layers are most effective for age and gender classification, and how does dimensionality reduction impact performance and efficiency?}
    We systematically identify the most informative layers for age and gender classification, demonstrating that dimensionality reduction preserves essential discriminative features while substantially enhancing classification accuracy.

\end{itemize}

\section{Proposed Framework}

\begin{figure}[!h]
    \centering
    \includegraphics[width=8cm]{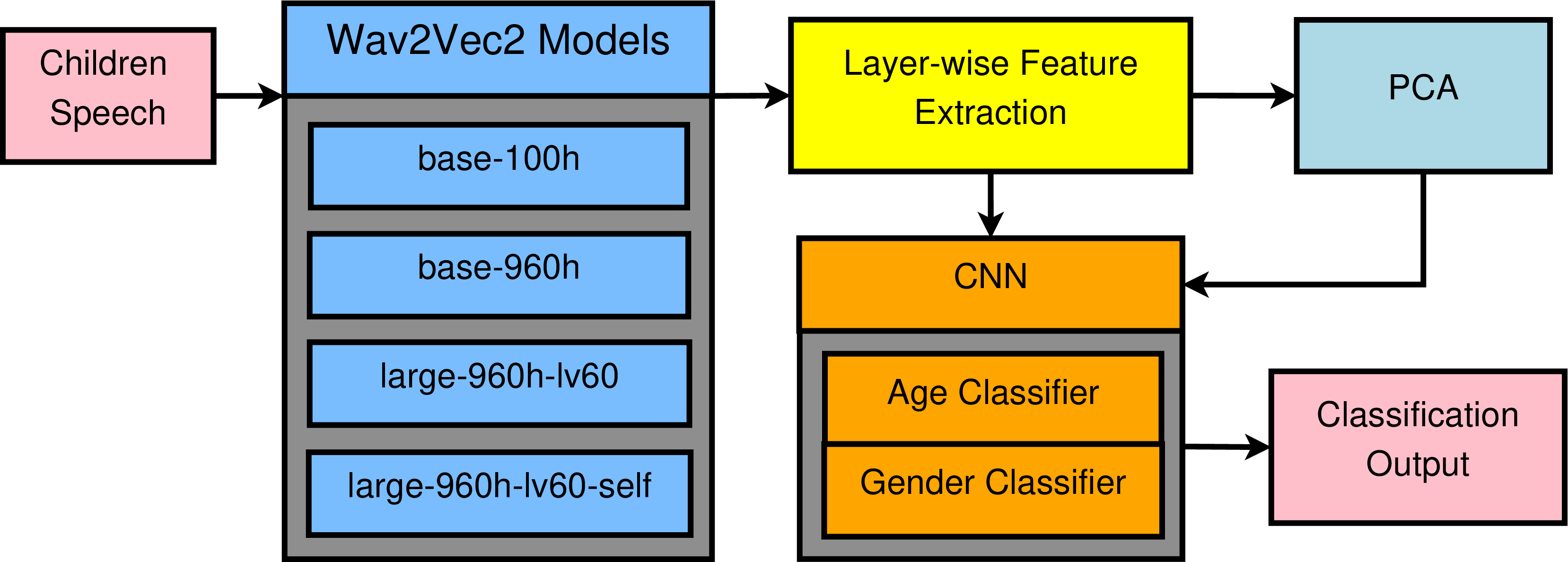} 
    \caption{Proposed framework for classifying age and gender in children using layer-wise features from various Wav2Vec2 models. Dimensionality reduction via PCA is applied to the most effective layers to enhance performance.}
    \label{fig:block}
    \vspace{-10pt}
\end{figure}

Figure \ref{fig:block} illustrates our framework for classifying age and gender in children’s speech using layer-wise features extracted from various Wav2Vec2 models. We evaluate four variants base-100h, base-960h, large-960h-lv60, and large-960h-lv60-self which differ in size, number of layers, and pre-training duration. Although these models are pre-trained on adult speech, we extract their layer-wise features without any fine-tuning, thereby assessing how well the learned representations generalize to children’s speech for age and gender classification.

The extracted features are fed into a convolutional neural network (CNN) to directly gauge the discriminative power of the Wav2Vec2 features without the influence of additional complexities. Moreover, we employ Principal Component Analysis (PCA) for dimensionality reduction to isolate and retain the most critical features, and thereby enhancing performance.

\section{Database and Experimental Setup}
\label{sec:format}

In this section, we describe the datasets and experimental setup used to evaluate SSL models on children’s speech. We explain the composition of the databases and outline the configurations used in our experiments.

\subsection{Database}

For this study, we use two widely recognized children’s speech datasets: PFSTAR \cite{russell2006pf} and CMU Kids \cite{eskenazi1997cmu}.

The PFSTAR dataset consists of recordings of children speaking British English, with ages ranging from 4 to 14 years. In our experiments, the PFSTAR training set consists of 8.3 hours of speech from 122 speakers, whereas the test set includes 1.1 hours of speech from 60 speakers.

\begin{table}[!h]
    \centering
    \vspace{-5pt}
    \caption{Dataset distribution for age and gender classification tasks using layer-wise features from Wav2Vec2 models for the PFSTAR and CMU Kids datasets.}
    \resizebox{8cm}{!}{
    \begin{tabular}{l c c c c c}
        \hline
        \multirow{2}{*}{\textbf{Dataset}} & \multirow{2}{*}{\shortstack{\textbf{Age} \\ \textbf{Range}}} & \multirow{2}{*}{\textbf{Split}} & \multirow{2}{*}{\shortstack{\textbf{Male} \\ \textbf{Speakers}}} & \multirow{2}{*}{\shortstack{\textbf{Female} \\ \textbf{Speakers}}} & \multirow{2}{*}{\shortstack{\textbf{Total} \\ \textbf{Utterances}}} \\
        & & & & & \\
        \hline
        \multirow{2}{*}{PFSTAR} & \multirow{2}{*}{4-14 yrs} & Train & 442 & 414 & 856 \\
                                &                         & Test  & 73  & 56  & 129 \\
        \hline
        \multirow{2}{*}{CMU Kids} & \multirow{2}{*}{6-11 yrs} & Train & 1217 & 2349 & 3566 \\
                                &                         & Test  & 529  & 1085 & 1614 \\
        \hline
    \end{tabular}}
    \label{tab:dataset_distribution}
    \vspace{-5pt}
\end{table}

The CMU Kids dataset contains recordings of children reading sentences in American English. This corpus includes 76 speakers aged between 6 and 11 years, with a total of 5,180 utterances. We use a 70/30 split for this dataset, allocating 6.3 hours for training and reserving 2.83 hours for testing.

Table \ref{tab:dataset_distribution} provides a comprehensive overview of the dataset distributions for age and gender classification. For the PFSTAR dataset, the distribution includes training and testing splits, covering the age range of 4 to 14 years (11 distinct age classes), with gender categories for male and female speakers. Similarly, the CMU Kids dataset is represented with its training and testing splits, covering the age range of 6 to 11 years (6 distinct age classes), along with gender statistics. Additionally, the figure highlights the total number of utterances for each dataset.

\begin{figure*}[!t]
    \centering
    \includegraphics[width=16.5cm, height=7.5cm]{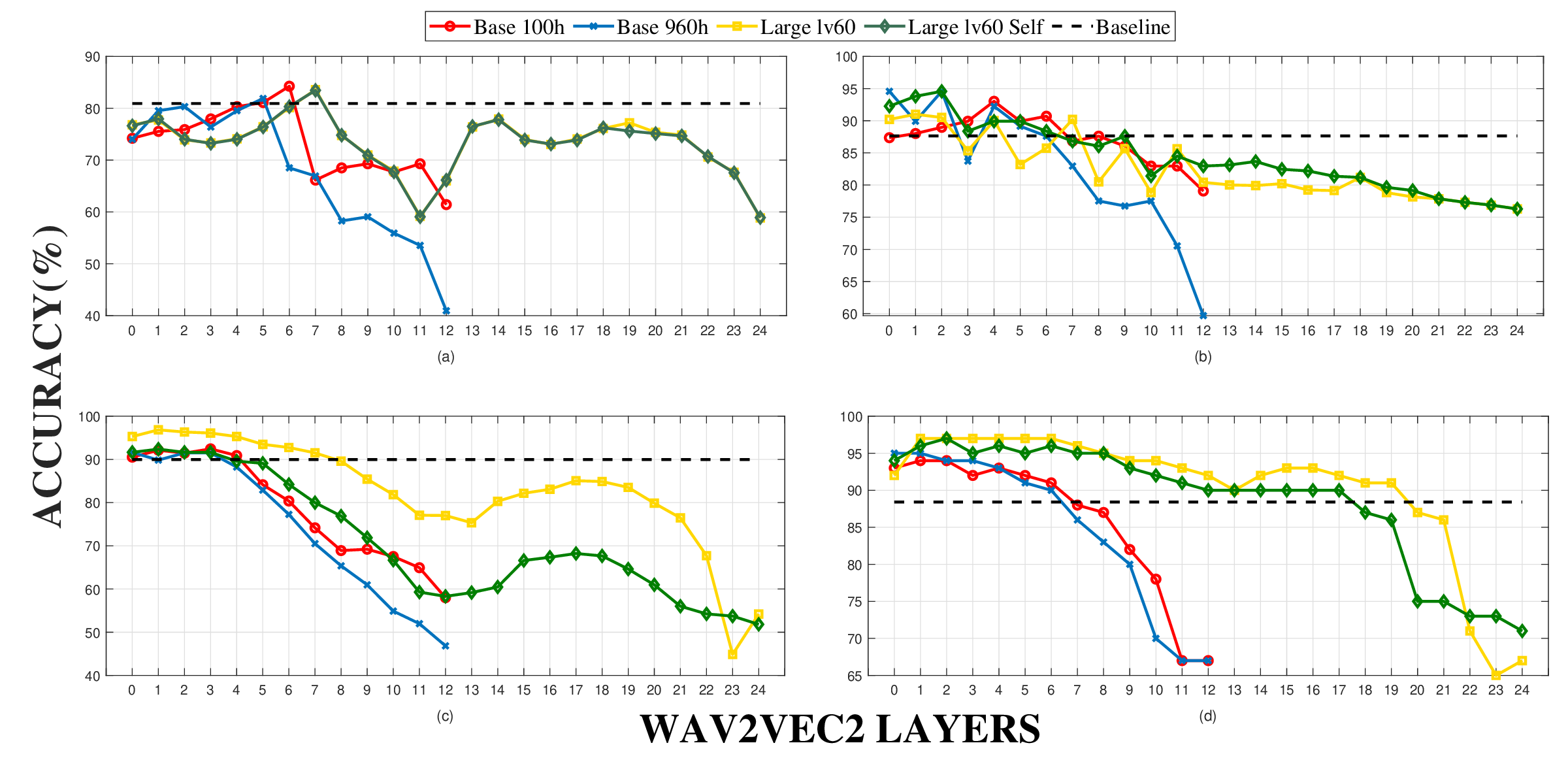}
    \caption{Layer-wise classification performance for age and gender across four Wav2Vec2 models. Subfigure (a) presents age classification results on the PFSTAR dataset; (b) shows gender classification on PFSTAR; (c) displays age classification on CMU Kids; and (d) illustrates gender classification on CMU Kids.}
    \label{fig:layer_wise}
    \vspace{-10pt}
\end{figure*}

\subsection{Experimental Setup}

For our experiments, we followed the framework illustrated in Figure \ref{fig:block}. In brief, we first extract layer-wise features from each Wav2Vec2 model and then input these features into a simple CNN architecture, consisting of three convolutional layers with 64, 128, and 256 filters, respectively, all using a kernel size of 5. We employ ReLU activations to introduce non-linearity, and batch normalization is applied to ensure stable training and enhance performance.

To evaluate Wav2Vec2 models for age and gender classification in children’s speech, we tested four variants:
\begin{itemize}
    \item \textbf{base-100h}: Trained on 100 hours of LibriSpeech data \cite{panayotov2015librispeech}, with fewer (95 million) parameters.

    \item \textbf{base-960h}: Trained on 960 hours of LibriSpeech data, capturing a broader range of features.

    \item \textbf{large-960h-lv60}: Trained on 960 hours of LibriSpeech and fine-tuned with 60,000 hours of unlabeled Libri-Light data \cite{kahn2020libri} to capture diverse speech patterns.

    \item \textbf{large-960h-lv60-self}: Enhanced further by self-training on the 960-hour LibriSpeech dataset.
\end{itemize}

\begin{table}[!h]
    \centering
    \caption{Wav2Vec2 model details. M is short of a million, K for a thousand and h for hours. PT represents pretraining, FT represents finetuning and Ft. Dim. represents feature dimensions for each model.}
    
    \resizebox{8cm}{!}{
    \begin{tabular}{|c|c|c|c|c|c|}
        \hline
        \textbf{Wav2Vec2} & \textbf{Size} & \textbf{PT} & \textbf{FT} & \textbf{Layers} & \textbf{Ft. Dim.}\\
        \hline
        base-100h & 95M & 960 h & 100 h & 13 & 768\\
        base-960h & 95M & 960 h & 960 h & 13 & 768\\
        large-960-lv60 & 317M & 60K h & 960 h & 25 & 1024\\
        large-960-lv60-self &317M & 60K h & 960 h &25 &1024\\
        \hline
    \end{tabular}%
    }
    \label{tab:wav2vec2_models}
    
\end{table}

The base models have 13 (0-12) hidden layers (768-dimensional features) with 95 million parameters and the large models have 25 (0-24) hidden layers (1024-dimensional features) with 317 million parameters. Each model first uses CNNs to transform raw audio into latent representations capturing local acoustic details. These are then refined by Transformer encoders using multiple self-attention and feed-forward layers to capture long-range dependencies. The initial hidden layer is the CNN output, followed by Transformer layers.

For the base models, this yields thirteen feature sets (one from the CNN output plus one for each Transformer layer), while the large models produce twenty-five sets. We then perform a comprehensive analysis of each layer's effectiveness in age and gender classification tasks. Finally, we apply PCA to the best-performing layers, identified from the layer-wise study, to evaluate the impact of dimensionality reduction on age and gender classification.

\section{Results and Discussion}
\label{sec: Result}

\subsection{Baseline}

We established a baseline using a CNN classifier trained on 26-dimensional Mel-Frequency Cepstral Coefficients (MFCCs) extracted from the PFSTAR and CMU Kids datasets. As shown in Tables~\ref{tab:PFSTAR} and~\ref{tab:CMU}, MFCCs achieved reasonable performance, particularly for gender classification, with accuracies of 87.63\% on PFSTAR and 88.41\% on CMU Kids. In contrast, age classification was more challenging, yielding 80.92\% and 89.97\% accuracy on PFSTAR and CMU Kids, respectively. The higher accuracy on CMU Kids is attributed to its smaller number of age classes (6 vs. 11 in PFSTAR), simplifying the classification task.

\subsection{Layer wise Wav2Vec2 performance}

To evaluate the effectiveness of SSL representations, we extracted layer-wise features from four Wav2Vec2 models: base-100h, base-960h, large-960h-lv60, and large-960h-lv60-self. Each feature set was passed through the same CNN classifier used for the baseline. Figure~\ref{fig:layer_wise} presents the classification accuracy across all layers for both age and gender tasks on the PFSTAR and CMU Kids datasets. The figure clearly demonstrates that early layers, which capture fundamental low level acoustic features, are essential for accurate age and gender classification, while later layers focused on more abstract representations are less effective for these specific tasks.

\begin{table}[!h]
    \centering
    \caption{Performance metrics- accuracy (A), precision (P), recall (R), and F1 score (F1) for the best performing layers across various Wav2Vec2 models in age and gender classification on the PFSTAR dataset (11 age classes).}
    \vspace{-10pt}
    \resizebox{8cm}{!}{
    \begin{tabular}{|c|c|c|c|c|c|} \hline 
        
        \multicolumn{6}{|c|}{\textbf{Age}} \\ \hline 
        
        \textbf{Wav2Vec2} & \textbf{Best} & \multirow{2}{*}{\textbf{A}} & \multirow{2}{*}{\textbf{P}} & \multirow{2}{*}{\textbf{R}} & \multirow{2}{*}{\textbf{F1}} \\   
        \textbf{Model} & \textbf{Layer} & & & & \\ \hline  
 Baseline (MFCC Features)& 26 Features& 80.92& 0.82& 0.81&0.80\\ \hline 
        
        base-100h & 6 & 84.25& 0.86& 0.84& 0.83                      \\ \hline 
        
        base-960h & 5 & 81.89& 0.84& 0.82& 0.81                      \\ \hline 
        
        large-960h-lv60 & 7 & 83.59& 0.84& 0.83& 0.83                      \\ \hline 
        
        large-960h-lv60-self & 7 & 83.46& 0.85& 0.84& 0.83                      \\ \hline 
        
        \multicolumn{6}{|c|}{\textbf{Gender}} \\ \hline  
 Baseline (MFCC Features)& 26 Features& 87.63& 0.90& 0.88&0.88\\ \hline 
        
        base-100h & 4 & 93.02& 0.93& 0.93& 0.92                      \\ \hline 
        
        base-960h & 2 & 94.57& 0.96& 0.95& 0.95                      \\ \hline 
        
        large-960h-lv60 & 1 & 91.45& 0.93& 0.92& 0.90                      \\ \hline 
        
        large-960h-lv60-self & 2 & 94.57& 0.95& 0.94& 0.94                      \\ \hline
        
    \end{tabular}
    }
    \label{tab:PFSTAR}
    \vspace{-10pt}
\end{table}

As summarized in Tables~\ref{tab:PFSTAR} and~\ref{tab:CMU}, the best-performing layers from Wav2Vec2 models significantly outperform the MFCC baseline for age classification across both datasets. Gender classification also improves in most models, though the gains are more pronounced in larger variants. On the PFSTAR dataset, the base-100h and large-960h-lv60 models deliver outstanding performance, achieving 84.25\% accuracy for age classification and 94.57\% accuracy for gender classification. For the CMU Kids dataset, the large-960h-lv60 model achieves exceptional results, with 96.84\% accuracy for age classification and 96.68\% for gender classification. 

\begin{table}[!b]
    \vspace{-15pt}
    \centering
    \caption{Performance metrics- accuracy (A), precision (P), recall (R), and F1 score (F1) for the best performing layers across various Wav2Vec2 models in age and gender classification on the CMU Kids dataset (6 age classes).}
    \vspace{-10pt}
    \resizebox{8cm}{!}{
    \begin{tabular}{|c|c|c|c|c|c|}
        \hline
        \multicolumn{6}{|c|}{\textbf{Age}} \\
        \hline
        \textbf{Wav2Vec2} & \textbf{Best} & \multirow{2}{*}{\textbf{A}} & \multirow{2}{*}{\textbf{P}} & \multirow{2}{*}{\textbf{R}} & \multirow{2}{*}{\textbf{F1}} \\
        \textbf{Model} & \textbf{Layer} & & & & \\ \hline 
 Baseline (MFCC Features)& 26 Features& 89.97& 0.90& 0.89&0.89\\
        \hline
        base-100h & 1 & 92.13& 0.92& 0.92& 0.92\\
        \hline
        base-960h & 0 & 91.63& 0.90& 0.90& 0.90\\
        \hline
        large-960h-lv60 & 1 & 96.84& 0.97& 0.97& 0.97\\
        \hline
        large-960h-lv60-self & 1 & 92.37& 0.92& 0.92& 0.92\\
        \hline
        \multicolumn{6}{|c|}{\textbf{Gender}} \\ \hline 
 Baseline (MFCC Features)& 26 Features& 88.41& 0.89& 0.88&0.88\\
        \hline
        base-100h & 2 & 93.78& 0.94& 0.94& 0.94\\
        \hline
        base-960h & 1 & 94.96& 0.95& 0.95& 0.95\\
        \hline
        large-960h-lv60 & 2 & 96.68& 0.97& 0.97& 0.97\\
        \hline
        large-960h-lv60-self & 2 & 96.53& 0.97& 0.97& 0.97\\
        \hline
    \end{tabular}
    }
    \vspace{-10pt}
    \label{tab:CMU}
\end{table}

\begin{figure*}[!t]
    \centering
    \includegraphics[width=17.2cm, height=7cm]{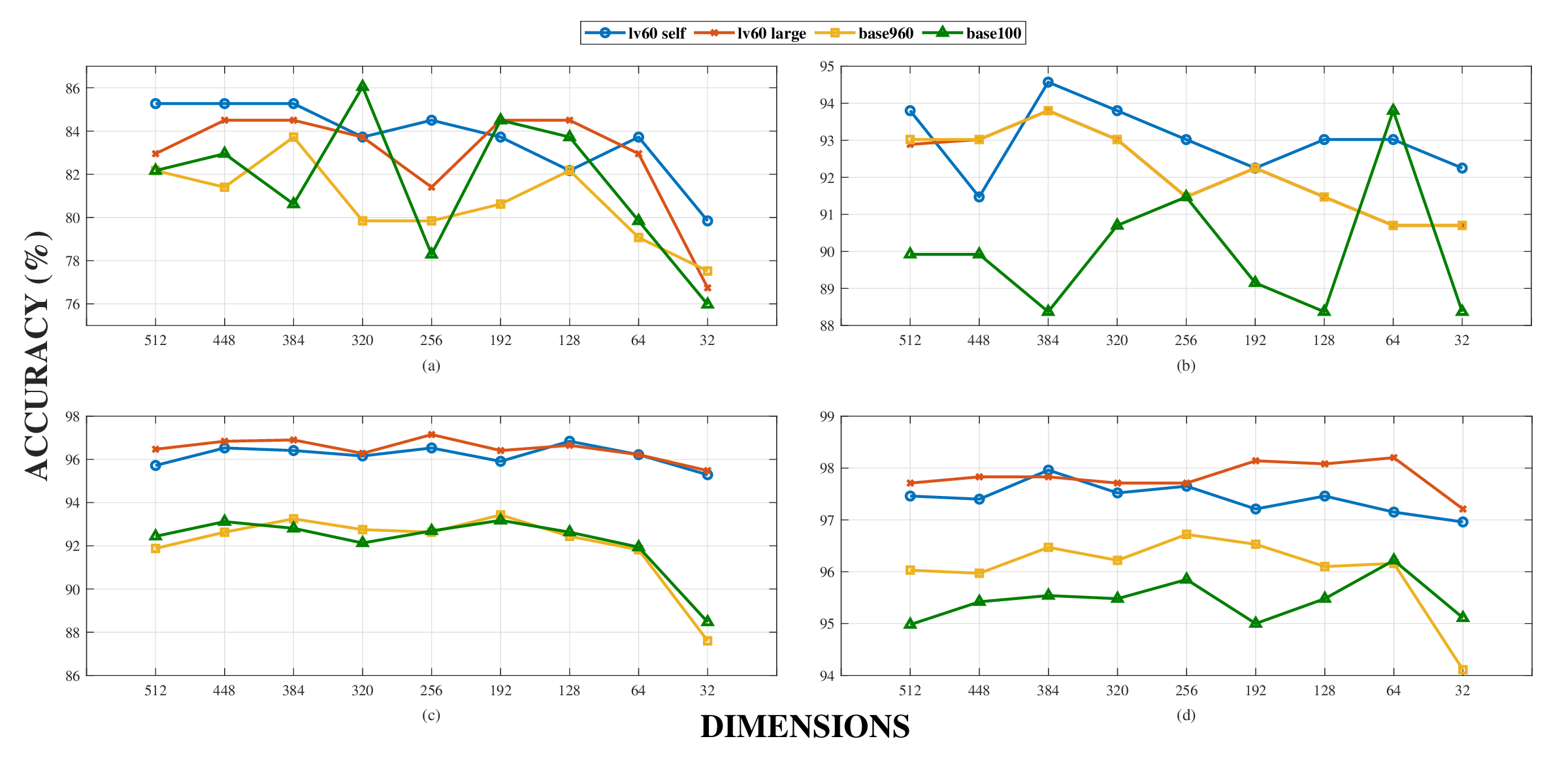}
    \caption{Age and gender classification performance with PCA-reduced features across four Wav2Vec2 models: Subfigure (a) presents age classification results on the PFSTAR dataset; (b) shows gender classification on PFSTAR; (c) displays age classification on CMU Kids; and (d) illustrates gender classification on CMU Kids.}
    \label{fig:pca}
    \vspace{-10pt}
\end{figure*}

To quantify these improvements, we performed Wilcoxon signed-rank tests comparing the MFCC baseline with the best-performing SSL layers. Results confirm statistically significant improvements for age classification across all models ($p < 0.001$). For gender, only larger models show significance, indicating that MFCCs already capture strong gender cues, limiting the margin for improvement with SSL features.

\subsection{Dimensionality Reduction using PCA}

We applied PCA to the best-performing layer from each Wav2Vec2 model to evaluate whether a reduced feature space could retain task-relevant information while improving computational efficiency. Dimensionality was progressively reduced from 512 to 32 in steps of 64, and classification accuracy was measured at each step.

As shown in Tables~\ref{tab:PCA_PFSTAR} and~\ref{tab:PCA_CMU} and illustrated in Figure~\ref{fig:pca}, PCA consistently preserved or even improved classification performance, particularly for age classification. On the PFSTAR dataset, the base-100h model achieved the highest age classification accuracy of 86.05\% with PCA reducing the feature dimensions to 320, while the large-960h-lv60-self model delivered the best gender classification performance with 95\% accuracy at 384 dimensions. On the CMU Kids dataset, the large-960h-lv60 model stood out, reaching 97.14\% accuracy for age classification and 98.20\% for gender classification with reduced feature dimensions.

\begin{table}[!b]
    \vspace{-10pt}
    \centering
    \caption{Performance metrics-accuracy (A), precision (P), recall (R), and F1 score (F1), for the best-performing layers of various Wav2Vec2 models in age and gender classification on the PFSTAR dataset (11 age classes), with reduced feature dimensions using PCA.}
    \vspace{-10pt}
    \resizebox{8cm}{!}{
    \begin{tabular}{|c|c|c|c|c|c|}
        \hline
        \multicolumn{6}{|c|}{\textbf{Age}} \\
        \hline
        \textbf{Wav2Vec2} & \textbf{Reduced Feature} & \multirow{2}{*}{\textbf{A}} & \multirow{2}{*}{\textbf{P}} & \multirow{2}{*}{\textbf{R}} & \multirow{2}{*}{\textbf{F1}} \\
        \textbf{Model} & \textbf{Dimension} & & & & \\
        \hline
        base-100h & 320 & 86.05& 0.87& 0.86& 0.86        \\
        \hline
        base-960h & 384 & 83.72& 0.85& 0.84& 0.82        \\
        \hline
        large-960h-lv60 & 256& 84.8& 0.85& 0.84& 0.83\\
        \hline
        large-960h-lv60-self & 384 & 85.27& 0.87& 0.85& 0.84        \\
        \hline
        \multicolumn{6}{|c|}{\textbf{Gender}} \\
        \hline
        base-100h & 64 & 93.80& 0.94& 0.93& 0.94        \\
        \hline
        base-960h & 384 & 93.80& 0.95& 0.94& 0.94        \\
        \hline
        large-960h-lv60 &320& 92.75& 0.94& 0.92& 0.91\\
        \hline
        large-960h-lv60-self &384 & 95.00& 0.95& 0.95& 0.95        \\
        \hline
    \end{tabular}
    }
    \label{tab:PCA_PFSTAR}
\end{table}

These results suggest that speaker-relevant information is highly concentrated in a subset of principal components. In particular, gender cues appear more compact and acoustically localized, while age-related traits require a slightly broader representation. Thus, dimensionality reduction not only reduces storage and computation but can also enhance generalization by removing redundant or noisy components.

\begin{table}[!t]
    \centering
    \caption{Performance metrics-accuracy (A), precision (P), recall (R), and F1 score (F1), for the best-performing layers of four Wav2Vec2 models in age and gender classification on the CMU Kids dataset (6 age classes), with reduced feature dimensions using PCA.}
    \vspace{-10pt}
    \resizebox{8cm}{!}{
    \begin{tabular}{|c|c|c|c|c|c|}
        \hline
        \multicolumn{6}{|c|}{\textbf{Age}} \\
        \hline
        \textbf{Wav2Vec2} & \textbf{Reduced Feature}& \multirow{2}{*}{\textbf{A}} & \multirow{2}{*}{\textbf{P}} & \multirow{2}{*}{\textbf{R}} & \multirow{2}{*}{\textbf{F1}} \\
        \textbf{Model} & \textbf{Dimension}& & & & \\
        \hline
        base-100h & 192                                & 93.18& 0.93& 0.93& 0.93        \\
        \hline
        base-960h & 192                                & 93.43& 0.93& 0.93& 0.93        \\
        \hline
        large-960h-lv60 & 256                                & 97.14& 0.97& 0.97& 0.97        \\
        \hline
        large-960h-lv60-self & 128                                & 96.84& 0.97& 0.97& 0.97        \\
        \hline
        \multicolumn{6}{|c|}{\textbf{Gender}} \\
        \hline
        base-100h & 64                                 & 96.22& 0.96& 0.96& 0.96        \\
        \hline
        base-960h & 256                                & 96.71& 0.97& 0.97& 0.97        \\
        \hline
        large-960h-lv60 & 64                                 & 98.20& 0.98& 0.98& 0.98        \\
        \hline
        large-960h-lv60-self & 384                                & 97.95& 0.98& 0.98& 0.98        \\
        \hline
    \end{tabular}
    }
    \vspace{-10pt}
    \label{tab:PCA_CMU}
\end{table}

\section{Conclusion}

In this work, we presented a comprehensive layer-wise analysis of four Wav2Vec2 variants for age and gender classification in children's speech. Our findings demonstrate that early layers capture speaker-specific attributes more effectively than deeper layers, which primarily encode linguistic information. This confirms the structured disentanglement of acoustic and linguistic representations across SSL model depth. By comparing with MFCC-based baselines, we established that self-supervised representations provide statistically significant improvements for age classification, while gender gains are more model-dependent due to the acoustic sufficiency of handcrafted features. Furthermore, our PCA-based dimensionality reduction experiments revealed that high classification accuracy can be achieved with compact feature sets, improving efficiency without sacrificing performance. These results offer new insights into how speaker traits are distributed within SSL models and highlight the potential for layer selection and feature compression in child-centric speech systems.

\section{Acknowledgements} This work was partially supported by the \textit{Simons Foundation}.

\bibliographystyle{IEEEtran}
\bibliography{mybib}

\end{document}